\documentstyle[eqsecnum,aps]{revtex}

\begin{document}
\title{Antihydrogen production and accuracy of the equivalent photon approximation 
\thanks{%
Dedicated to the memory of Franz Osterfeld.}}
\author{C.A. Bertulani$^{(a)}$ and G. Baur$^{(b)}$}
\address{$^b$Instituto de F\'\i sica,
Universidade Federal do Rio de Janeiro, 
21945-970 Rio de Janeiro, RJ, Brazil, E-mail: bertu@if.ufrj.br}
\address{$^b$ Forschungszentrum J\"ulich, Postfach 1913, 
D-52425 J\"ulich, Germany, E-mail: G.Baur@fz-juelich.de}
\date{\today }
\maketitle

\begin{abstract}
The production of antihydrogen in flight in $\overline{{\rm p}}$-nucleus
collisions is calculated theoretically in the Plane Wave Born Approximation
(which is equivalent to the straight line semiclassical approximation).
Antihydrogen has been produced in this way at LEAR/CERN \cite{Ba96} and is
presently studied at FERMILAB at various $\overline{{\rm p}}$ energies.
Dirac wave functions for the leptons are used, taking first order ($Z\alpha $%
) corrections into account. Analytical results are obtained for differential
cross-sections. Total cross sections are obtained by numerical integration.
The dependence on the transverse momentum transfer is studied and the
accuracy of the equivalent photon approximation and a recent variant by
Munger, Brodsky,and Schmidt \cite{Mu94} is discussed as a function of beam
energy.
\end{abstract}

\bigskip


\bigskip


\section{Introduction}

Antihydrogen in flight was first produced and detected at CERN/LEAR \cite
{Ba96} using the process

\begin{equation}
\overline{{\rm p}}+Z\longrightarrow \overline{{\rm H}}+e^{-}+Z\;,
\end{equation}
where Z=54 (Xe). This process was first suggested by Munger et al. \cite
{Mu94},where also calculations of the cross section for the antihydrogen
production were performed in the equivalent photon approximation (EPA). A
similar pair production process with e-capture will occur with large cross
section at relativistic heavy ion colliders \cite{BB94}. This is important
because it leads to a beam loss. Theoretical calculations were done by many
groups \cite{Be88,As94,Ba93}. They were mainly interested in the limit of
high energies. The applicability of EPA will become more and more
questionable for the lower energies, where the $\overline{{\rm H}}$
production experiments were done \cite{Ba96,Bau93}. The dominant graph is
shown in figure 1. It was shown in \cite{Bau93} that bremsstrahlung pair
production with capture can be safely neglected.

By crossing symmetry the process $\gamma +\overline{{\rm p}}\longrightarrow
e^{-}+\overline{{\rm H}}$ \cite{Mu94} is related to the photoeffect $\gamma
+ $ ${\rm H}\longrightarrow {\rm p}+e^{-}$ \cite{Be88} which is treated in
the literature (see e.g.\cite{Be79}). These calculations provide approximate
theoretical values for the $\overline{{\rm H}}$ production cross section.
This is also relevant for the ongoing experiments at Fermilab \cite{Fe97},
where also the energy dependence of the process will be investigated and
possible future experiments with fast $\overline{{\rm H}}$ beams, like Lamb
shift measurements \cite{Mu94,Fe97}. On the other hand, we have to
investigate the accuracy and limitations of the EPA. It is the purpose of
this paper to carry out an accurate calculation of the cross section for eq.
(1.1). Analytical expressions are given.

\section{Production of $\overline{{\rm H}}$ with antiproton beams}

In the semiclassical approach we assume that the relativistic antiproton
moves along a straight line. In its frame of reference the time-dependent
electromagnetic fields of the target nucleus is given by

\begin{equation}
{\bf A}({\bf r},t)={\bf v\;}\phi ({\bf r},t),{\rm \ \ \ where\ \ }\phi ({\bf %
r},t)=\frac{Z_Te}{\left| {\bf r-r}^{\prime }(t)\right| }\;,
\end{equation}
where ${\bf r}=(x,y,z)$, and ${\bf r}^{\prime }=(b_x,b_y,\gamma {\rm v}t)$,
where ${\rm v}$ is the relative velocity, $\gamma =(1-{\rm v}^2)^{-1/2}$ ,
and $Z_Te$ is the target charge. We can write $\phi $ in the integral
representation

\begin{equation}
\phi ({\bf r},t)=\frac{Z_Te}{2\pi ^2}\int d^3q\;\frac{e^{i{\bf q.}\left[ 
{\bf r-r}^{\prime }(t)\right] }}{q^2}\;.
\end{equation}

In the first-order perturbation theory, the $\overline{{\rm H}}$-production
amplitude for a collision with impact parameter $b=\sqrt{b_x^2+b_y^2}$ is
given in terms of the transition density and current, $\rho \left( {\bf r}%
\right) $ and ${\bf j(r),}$respectively, by

\begin{eqnarray}
a_{1st} &=&\frac 1i\int dt\;e^{i\omega t}\int d^3r\;\left[ \rho \left( {\bf r%
}\right) \phi ({\bf r},t)-{\bf j(r).A}({\bf r},t)\right]  \nonumber \\
&=&\frac{Z_Te}{i\pi {\rm v}}\int dt\;e^{i\omega t}\int d^3q\int d^3r\;\frac{%
e^{i{\bf q.}\left[ {\bf r-r}^{\prime }(t)\right] }}{q^2}\left[ \rho \left( 
{\bf r}\right) -{\bf v.j}({\bf r})\right] \;,
\end{eqnarray}
where $\omega =\varepsilon +m$ is the sum of the electron, $\varepsilon $,
and the positron energy, $m$ (the binding energy of $\overline{{\rm H}}$ can
be neglected). In the last step we have used eqs. (2) and (3). Performing
the integral over time and using the continuity equation, $\nabla .{\bf j=-}%
i\omega \rho $, we get

\begin{equation}
a_{1st}=\frac{Z_Te}{i\pi {\rm v}}\int d^2q_t\frac{e^{i{\bf q}_t{\bf .b}}}{Q^2%
}\int d^3r\;e^{i{\bf Q.r}}\;\left( \frac{j_z({\bf r})}{{\rm v}\gamma ^2}+%
\frac{{\bf q}_t.{\bf j}_t}\omega \right) ,
\end{equation}
where the index $z$ $(t)$ denotes the components along (transverse to) the
beam direction, and ${\bf Q}\equiv ({\bf q}_t,\omega /{\rm v}).\;$The
transition current is given in terms of the Dirac matrices, $\overrightarrow{%
\alpha },$and the lepton wavefunctions ${\bf j}(r)=e\left[ \Psi
^{(-)}\right] ^{*}\overrightarrow{\alpha }\Psi ^{(+)}.$ Thus we can rewrite
the $\overline{{\rm H}}$-production as

\begin{equation}
a_{1st}=\frac{Z_Te}{i\pi {\rm v}}\int d^2q_t\;\frac{e^{i{\bf q}_t{\bf .b}}}{%
\left[ q_t^2+\left( \omega /\gamma {\rm v}\right) ^2\right] ^2}\;F({\bf Q}%
)\;,
\end{equation}
where

\begin{eqnarray}
F({\bf Q}) &=&e\int d^3r\;\left[ \Psi ^{(-)}\right] ^{*}({\bf r})\;e^{i{\bf %
Q.r}}\;\left( \frac{\alpha _z}{{\rm v}\gamma ^2}+\frac{{\bf q}_t.%
\overrightarrow{\alpha }_t}\omega \right) \Psi ^{(+)}({\bf r})  \nonumber \\
&=&ie\;{\bf K\;.}\int d^3r\;\left[ \Psi ^{(-)}\right] ^{*}({\bf r})\;e^{i%
{\bf Q.r}}\;\gamma ^0\overrightarrow{\gamma }{\bf \;}\Psi ^{(+)}({\bf r})\;.
\end{eqnarray}
In the last equality we used the definition ${\bf K}\equiv \left( {\bf K}%
_t,K_z\right) =\left( {\bf q}_t/\omega ,1/{\rm v}\gamma ^2\right) $ and ($%
\gamma ^0,\overrightarrow{\gamma }{\bf )}$ are the Dirac matrices, where we
follow the procedure of ref. \cite{Be79}. The total cross section is
obtained by integrating the square of the expression (8) over all possible
impact parameters:

\begin{equation}
\sigma =\sum_{spins}\int \;\left| a_{1st}\right| ^2d^2b=4\left( \frac{Z_Te}{%
{\rm v}}\right) ^2\sum_{spins}\int_0^\infty d^2q_t\;\frac{\left| F({\bf Q}%
)\right| ^2}{(Q^2-\omega^2)^2}\;.
\end{equation}
The same result can be obtained in the Plane-Wave Born approximation, as
shown in the appendix C.

Using the positron and the electron wavefunctions as given by eqs. (A.4) of
appendix A and eq. (B.1) of the appendix B, we get

\begin{eqnarray}
F({\bf Q}) &=&ie\int d^3r\;\left\{ \overline{u}\;e^{-i{\bf p.r}}+\overline{%
\Psi ^{\prime }}\right\} \left( \overrightarrow{\gamma }{\bf .K}\right)
\;e^{i{\bf Q.r}}\;  \nonumber \\
\times &&\left\{ 1+\frac i{2m}\;\gamma ^0\overrightarrow{\gamma }{\bf .}%
\overrightarrow{\nabla }\right\} \;{\rm v}\Psi _{non-r}({\bf r})  \nonumber
\\
&\simeq &ie\int d^3r\;\Bigg\{ \overline{u}\;e^{i({\bf Q}-{\bf p).r}}\left( 
\overrightarrow{\gamma }{\bf .K}\right) \;\left[ \left( 1+\frac i{2m}%
\;\gamma ^0\overrightarrow{\gamma }{\bf .}\overrightarrow{\nabla }\right)
\right] {\rm v}\Psi _{non-r}({\bf r})\;  \nonumber \\
+ &&\overline{\Psi ^{\prime }}\;e^{i{\bf Q.r}}\;\left( \overrightarrow{%
\gamma }{\bf .K}\right) {\rm v}\Psi _{non-r}({\bf r})\Bigg\}\;,
\end{eqnarray}
where in the last equation we neglected terms of highest order in $Z\alpha .$

Integrating by parts

\begin{eqnarray}
F({\bf Q}) &=&ie\int d^3r\;\{\overline{u}\;e^{i({\bf Q}-{\bf p).r}}\left( 
\overrightarrow{\gamma }{\bf .K}\right) \;\left[ \left( 1+\frac 1{2m}%
\;\gamma ^0\overrightarrow{\gamma }{\bf .}\overrightarrow{\nabla }\right) 
{\rm v}\Psi _{non-r}({\bf r})\right] \;  \nonumber \\
+ &&\overline{\Psi ^{\prime }}\;e^{i{\bf Q.r}}\left( \overrightarrow{\gamma }%
{\bf .K}\right) {\rm v}\Psi _{non-r}({\bf r})\}  \nonumber \\
&\simeq &\frac{ie}{\sqrt{\pi }}\;a_0^{-3/2}\int d^3r\;\Bigg\{\overline{u}%
\;\left( \overrightarrow{\gamma }{\bf .K}\right) \;  \nonumber \\
&&\left[ \left( 1+\frac 1{2m}\;\gamma ^0\overrightarrow{\gamma }{\bf .(Q-p)}%
\right) {\rm v\;}\left( e^{-r/a_0}\right) _{{\bf Q-p}}\right] \;+\overline{%
\Psi _{{\bf Q}}^{\prime }}\left( \overrightarrow{\gamma }{\bf .K}\right) \;%
{\rm v}\Bigg\}\;.  \nonumber \\
&&
\end{eqnarray}

Since the corrections are in first order in $Z\alpha $, we have replaced $%
\Psi _{non-r}({\bf r})$ by its constant value 1/$\sqrt{\pi }a_0^{3/2}$ in
the last term of eq. (2.9). If we would do the same in the first term it
would be identically zero (except for ${\bf Q=p}$). This is the reason why
we need to keep the corrections in the wavefunctions to first order in $%
Z\alpha $; for $\varepsilon \gg m$ these corrections yield a term of the
same order as the plane-wave to the total cross section, as we shall see
later. We rewrite eq. (2.9) as

\begin{equation}
F({\bf Q})=4ie\sqrt{\pi }a_0^{-5/2}\;\frac{\overline{{\rm u}}A{\rm v}}{%
\left( {\bf Q-p}\right) ^2}\;,
\end{equation}
where

\begin{equation}
A=a(\overrightarrow{\gamma }{\bf .K})+(\overrightarrow{\gamma }{\bf .K}%
)\gamma ^0\left( \overrightarrow{\gamma }{\bf .b}\right) +(\overrightarrow{%
\gamma }{\bf .d})\gamma ^0(\overrightarrow{\gamma }{\bf .K})\;,
\end{equation}
with

\begin{eqnarray}
a &=&\frac 1{\left( {\bf Q-p}\right) ^2}-\frac \varepsilon {Q^2-p^2}%
\;,\;\;\;\;{\bf b}=\frac{{\bf Q-p}}{2m\left( {\bf Q-p}\right) ^2}\;,\;\;\;\;%
{\rm and}  \nonumber \\
{\bf c} &=&-\frac{{\bf Q-p}}{2m(Q^2-p^2)}\;.
\end{eqnarray}
Inserting (2.10) into (2.7), the cross section becomes

\begin{equation}
\sigma =\frac{16e^2}{\pi ^2a_0^5}\;\left( \frac{Z_Te}{{\rm v}}\right) ^2\int 
\frac{d^3p}{\left( 2\pi \right) ^3}\int_0^\infty \frac{d^2q_t}{%
(Q^2-\omega^2)^2}\;\frac 1{\left( {\bf Q-p}\right) ^4}\sum_{spins}\left| 
\overline{{\rm u}}A{\rm v}\right| ^2.
\end{equation}

The sum runs over the spins of the positron and the electron. Using $%
p^2dp=\varepsilon pd\varepsilon $, and standard trace techniques we get

\begin{eqnarray}
\frac{d\sigma }{d\varepsilon d\Omega } &=&\frac{32e^2p}{\pi ^2a_0^5}\;\left( 
\frac{Z_Te}{{\rm v}}\right) ^2\int_0^\infty \frac{d^2q_t}{(Q^2-\omega^2)^2}\;%
\frac 1{\left( {\bf Q-p}\right) ^4}  \nonumber \\
&\times &\Bigg\{\left( \varepsilon -1\right) \left[ \left( {\bf b-d}\right)
^2K^2+4\left( {\bf K.b}\right) \left( {\bf K.d}\right) \right]  \nonumber \\
+ &&\left( \varepsilon +1\right) a^2K^2+2a\left[ 2\left( {\bf p.K}\right)
\left( {\bf b.K}\right) -{\bf p}.({\bf b-d})K^2\right] \Bigg\}
\end{eqnarray}

Using the definition of ${\bf K}$, it is illustrative to separate the
integrand of the above equation in terms of longitudinal and transverse
components:

\begin{equation}
\frac{d\sigma }{d\varepsilon d\Omega }=\frac 1{\pi \omega }\;\left( \frac{%
Z_Te}{{\rm v}}\right) ^2\int_0^\infty d(q_t^2)\;\frac{q_t^2}{(Q^2-\omega^2)^2%
}\;\Bigg[\frac{d\sigma _{\gamma ^{*}T}}{d\Omega }\left( {\bf Q},\omega
\right) +\frac{d\sigma _{\gamma ^{*}L}}{d\Omega }\left( {\bf Q},\omega
\right) +\frac{d\sigma _{\gamma ^{*}LT}}{d\Omega }\left( {\bf Q},\omega
\right) \Bigg]\;,
\end{equation}

where

\begin{eqnarray}
\frac{d\sigma _{\gamma ^{*}T}}{d\Omega }\left( {\bf Q},\omega \right) = &&%
\frac{32e^2p}{a_0^5\omega }\frac 1{\left( {\bf Q-p}\right) ^4}\;\Bigg\{%
\left( \varepsilon -1\right) \left[ \left( {\bf b-d}\right) ^2-\frac{\left(
q_t-{\bf p.}\widehat{{\bf e}}_t\right) {\bf p.}\widehat{{\bf e}}_t}{\left( 
{\bf Q-p}\right) ^2\left( Q^2-p^2\right) }\right] \\
+ &&\left( \varepsilon +1\right) a^2+2a\left[ \frac{\left( q_t-{\bf p.}%
\widehat{{\bf e}}_t\right) {\bf p.}\widehat{{\bf e}}_t}{\left( {\bf Q-p}%
\right) ^2}-{\bf p}.({\bf b-d})\right] \Bigg\}\;,  \nonumber
\end{eqnarray}

\begin{eqnarray}
\frac{d\sigma _{\gamma ^{*}L}}{d\Omega }\left( {\bf Q},\omega \right) &=&%
\frac{32e^2p\omega }{a_0^5q_t^2}\left( \frac 1{\gamma ^2{\rm v}}\right) ^2%
\frac 1{\left( {\bf Q-p}\right) ^4}\;\Bigg\{\left( \varepsilon -1\right)
\left[ \left( {\bf b-d}\right) ^2-\frac{\left( \omega /{\rm v}-p_z\right) ^2%
}{\left( {\bf Q-p}\right) ^2\left( Q^2-p^2\right) }\right]  \nonumber \\
+ &&\left( \varepsilon +1\right) a^2+2a\left[ \frac{p_z\left( \omega /{\rm v}%
-p_z\right) }{\left( {\bf Q-p}\right) ^2}-{\bf p}.({\bf b-d})\right] \Bigg\}%
\;,  \nonumber
\end{eqnarray}

\begin{eqnarray}
\frac{d\sigma _{\gamma ^{*}LT}}{d\Omega }\left( {\bf Q},\omega \right) &=&%
\frac{64e^2p}{a_0^5q_t\gamma ^2{\rm v}}\frac{\left( \omega /{\rm v}%
-p_z\right) }{\left( {\bf Q-p}\right) ^6} \\
&&\times \left\{ a\left[ \left( {\bf p.}\widehat{{\bf e}}_t\right) +p_z\frac{%
\left( q_t-{\bf p.}\widehat{{\bf e}}_t\right) }{\left( \omega /{\rm v}%
-p_z\right) }\right] -\left( \varepsilon -1\right) \frac{\left( q_t-{\bf p.}%
\widehat{{\bf e}}_t\right) }{\left( Q^2-p^2\right) }\right\} \;,  \nonumber
\end{eqnarray}

In these equations $\widehat{{\bf e}}=\widehat{{\bf q}}_t/q_t$ is a unit
vector in the transverse direction. The cross sections $\sigma _{\gamma
^{*}T}$, $\sigma _{\gamma ^{*}L}$, and $\sigma _{\gamma ^{*}LT}$ are
interpreted as the cross sections for $\overline{{\rm H}}$ by virtual
transverse and longitudinal photons and a interference term, respectively.
Note that the transverse and longitudinal directions are {\it with respect
to the beam axis},nor with respect to the photon momentum, as is usually
meant by this term. Only for $\gamma \gg 1$, this definition agrees with the
definition of transverse and longitudinal virtual photons. However, this
separation is very useful, as we will see next.

\section{Production of $\overline{{\rm {H}}}$ by real photons}

The cross section for production of $\overline{{\rm H}}$ by real photons is
given by \cite{Be79}

\begin{equation}
d\sigma _\gamma =2\pi \left| V_{fi}\right| ^2\delta (\omega -\varepsilon -1)%
\frac{d^3p}{\left( 2\pi \right) ^3}\;,
\end{equation}
where

\begin{equation}
V_{fi}=-e\sqrt{\frac{4\pi }\omega }%
\displaystyle \int 
d^3r\;\left[ \Psi ^{(-)}({\bf r})\right] ^{*}\;e^{i{\bf \kappa .r}}\;\left( 
\widehat{{\bf e}}.\overrightarrow{\alpha }\right) \Psi ^{(+)}({\bf r})\;,
\end{equation}
with ${\bf \kappa }=\widehat{{\bf z}}\;\omega /c$, where $\widehat{{\bf z}}$
is the unit vector along the photon incident direction, and $\widehat{{\bf e}%
}$ is photon polarization unit vector.

Using the positron and the electron wavefunctions as given by eqs. (A.4) of
appendix A and eq. (B.1) of the appendix B, and performing similar steps as
in section 2, we get

\begin{equation}
V_{fi}=-e\frac{8\pi \sqrt{2}}{a_0^{5/2}\omega ^{1/2}}\frac{uB{\rm v}}{\left( 
{\bf \kappa -p}\right) ^2}\;,
\end{equation}
where

\begin{equation}
B=a_\kappa (\overrightarrow{\gamma }{\bf .}\overrightarrow{{\bf \kappa }})+(%
\overrightarrow{\gamma }{\bf .}\overrightarrow{{\bf \kappa }})\gamma
^0\left( \overrightarrow{\gamma }{\bf .b}_\kappa \right) +(\overrightarrow{%
\gamma }{\bf .d}_\kappa )\gamma ^0(\overrightarrow{\gamma }{\bf .}%
\overrightarrow{{\bf \kappa }})\;.
\end{equation}
where $a_\kappa $, ${\bf b}_\kappa $, and ${\bf d}_\kappa $ are the
quantities defined in eqs. (2.12), but with ${\bf Q}$ replaced by $%
\overrightarrow{{\bf \kappa }}=\widehat{{\bf z}}\;\omega /c$. Inserting
these results in eq. (3.1) and summing over spins we get

\begin{eqnarray}
\frac{d\sigma _\gamma }{d\Omega }\left( \omega \right) &=&\frac{32e^2p}{%
a_0^5\omega }\frac 1{\left( {\bf \kappa -p}\right) ^4}  \nonumber \\
\times &&\Bigg\{\left( \varepsilon -1\right) \left[ \left( {\bf b}_\kappa 
{\bf -d}_\kappa \right) ^2-\frac{\left( {\bf p.}\widehat{{\bf e}}\right) ^2}{%
\left( {\bf \kappa -p}\right) ^2\left( \kappa ^2-p^2\right) }\right] 
\nonumber \\
+ &&\left( \varepsilon +1\right) a_\kappa ^2+2a_\kappa \left[ \frac{\left( 
{\bf p.}\widehat{{\bf e}}\right) ^2}{\left( {\bf \kappa -p}\right) ^2}-{\bf p%
}.({\bf b}_\kappa {\bf -d}_\kappa )\right] \Bigg\}\;.
\end{eqnarray}

We notice that the above equation can also be obtained from eqs. (2.15-2.18)
in the limit $q_t\rightarrow 0$ and ${\rm v}\longrightarrow c$. In this
limit $\sigma _{\gamma ^{*}L}\longrightarrow 0$, and $\sigma _{\gamma
^{*}LT}\longrightarrow 0$, and $\sigma _{\gamma ^{*}T}$ becomes the cross
section for the production $\overline{{\rm H}}$ by real photons.

Integrating eq. (3.5) over the azimuthal angle we get (without any further
approximations!)

\begin{equation}
\frac{d\sigma _\gamma }{d\theta }\left( \omega \right) =2\pi \frac{%
e^2p^3\left( \varepsilon +2\right) }{a_0^5\omega ^4}\frac{\sin ^3\theta }{%
\left( \varepsilon -p\cos \theta \right) ^4}\left[ \varepsilon -p\cos \theta
-\frac 2{\omega (\varepsilon +2)}\right]
\end{equation}

In figure 2 we plot the cross section for the angular distribution of the
electrons in $\overline{{\rm {H}}}$ production by real photons. We observe
that the higher the electron energy is, the more forward peaked the
distribution becomes. As we can immediately deduce from eq. (3.6), for $%
\varepsilon \gg 1$, the width of the peak is given by $\Delta \cos \theta
\simeq 1.$ For $\varepsilon \simeq 1$ the distribution is proportional to $%
\sin ^2\theta $.

Integrating eq. (3.6) over $\theta $ we get

\begin{equation}
\sigma \left( \gamma +\overline{{\rm {p}}}\longrightarrow e^{-}+\overline{%
{\rm {H}}}\right) =\frac{4\pi e^2}{a_0^5}\;\frac p{\omega ^4}\;\left[
\varepsilon ^2+\frac 23\varepsilon +\frac 43-\frac{\varepsilon +2}p\ln
\left( \varepsilon +p\right) \right]
\end{equation}

It is instructive to derive this cross section by a different method. In
some textbooks (see, e.g., ref. \cite{He54,Be79}) one can find a calculation
of the cross sections for the annihilation of a positron with an electron in
the K-shell of a nucleus. From charge conjugation invariance (which is valid
for QED), this cross section is the same as the cross section for the
annihilation of an electron with a positron in the K-shell of an $\overline{%
{\rm {H}}}$. The cross section is\ \cite{He54,err1}

\begin{equation}
\sigma \left( e^{-}+\overline{{\rm {H}}}\longrightarrow \gamma +\overline{%
{\rm {p}}}\right) =\frac{2\pi e^2}{a_0^5}\;\frac 1{p\omega ^2}\;\left[
\varepsilon ^2+\frac 23\varepsilon +\frac 43-\frac{\varepsilon +2}p\ln
\left( \varepsilon +p\right) \right]
\end{equation}

Using the detailed balance theorem, the above equation yields exactly the
same cross section of $\overline{{\rm {H}}}$ production by real photons as
given by eq. (3.7). The reason is that the calculation leading to eq. (3.7),
originally due to Sauter \cite{Sau}, used the same corrections to $Z\alpha $
order for the electron and positron wavefunctions, as we did to obtain eq.
(3.1) (see Appendices A and B).

The relevance of these corrections can be better understood by using the
high energy limit, $\varepsilon \gg 1.$ From eq. (3.7) we get

\begin{equation}
\sigma \left( \gamma +\overline{{\rm {p}}}\longrightarrow e^{-}+\overline{%
{\rm {H}}}\right) =\frac{4\pi e^2}{a_0^5}\;\frac 1\varepsilon \;
\end{equation}
If in our calculation, we had used plane-waves for the electron and the
hydrogenic wave function for the positron, without the corrections to order $%
Z\alpha $, the cross section for the production of $\overline{{\rm {H}}}$ by
real photons would be given by

\begin{equation}
\left[ \frac{d\sigma _\gamma }{d\theta }\left( \omega \right) \right]
_{PWA}=4\pi \frac{e^2p}{a_0^5\omega ^4}\frac{\sin \theta }{\left(
\varepsilon -p\cos \theta \right) ^4}\;,
\end{equation}
the integral of which being

\begin{equation}
\left[ \sigma _\gamma \right] _{PWA}=\frac{8\pi e^2}{a_0^5}\;\frac p{\omega
^4}\;\left( 3\varepsilon ^2+p^2\right) \;.
\end{equation}
In the high energy limit:

\begin{equation}
\left[ \sigma _\gamma \right] _{PWA}=\frac{32\pi e^2}{3a_0^5}\;\frac 1%
\varepsilon
\end{equation}

We see that the difference between this result and that of eq. (3.9) is a
factor 8/3. This is just the reason why the corrections of order $Z\alpha $
have to be included in the calculation of $\overline{{\rm {H}}}.$ They yield
terms to the matrix elements for photo-production of $\overline{{\rm {H}}}$
of the same order as the terms of lowest order. The origin of the difference
between the two results are the small distances which enter in the
calculation of the matrix elements of eq. (2.9). The corrections are
essential to account for their effects properly. We note that these
corrections are enough to account for a good description of $\overline{{\rm {%
H}}}$ production. 

The connection between the production of $\overline{{\rm {H}}}$ by real
photons and by virtual ones will prove to be very useful, as we shall see in
the next section.

\section{ Production of $\overline{{\rm {H}}}$ by virtual photons}

As we have seen in last section, the production of $\overline{{\rm {H}}}$ in
a collision of $\overline{{\rm {p}}}$ with nuclear targets is directly
related to the production cross section by real photons in the limit $%
q_t\rightarrow 0$ and ${\rm v}\longrightarrow c$. It is important to
determine at which values of $q_t$ and $\gamma $ this condition applies. In
figures 3 and 4 we plot the production cross sections by longitudinal and
transverse virtual photons, $\sigma _{\gamma ^{*}T}$, $\sigma _{\gamma
^{*}L} $, respectively. These are the integrals of eqs. (2.16) and (2.17),
which are performed numerically. The integral of the interference term, $%
\sigma _{\gamma ^{*}LT}$, eq. (2.18), is very small and is not shown. In
figure 3 the cross sections are calculated for $\varepsilon =2,$ and $\gamma
=3$ $,$ and in figure 4 for $\gamma =10$ and $\varepsilon =2$. We observe
that for low $\gamma $'s the cross section $\sigma _{\gamma ^{*}L}$, is as
important as $\sigma _{\gamma ^{*}T}$, while for large $\gamma $'s the cross
section $\sigma _{\gamma ^{*}T}$ dominates. For large $\gamma $'s the cross
section $\sigma _{\gamma ^{*}L}$ is only relevant at $q_t\simeq 0$. But this
region contributes little to the total cross section, which is given by

\begin{equation}
\sigma =\frac 1{\pi \omega }\;\left( \frac{Z_Te}{{\rm v}}\right)
^2\int_1^\infty d\varepsilon \;\int_0^\infty d(q_t^2)\;\frac{\;q_t^2}{%
(Q^2-\omega ^2)^2}\;\left[ \sigma _{\gamma ^{*}T}\left( Q,\omega \right)
+\sigma _{\gamma ^{*}L}\left( Q,\omega \right) +\sigma _{\gamma ^{*}L}\left(
Q,\omega \right) \right] \;,
\end{equation}
since the first term inside the integral suppresses $q_t\simeq 0$ values.
This can also be seen by calculating the total cross section for $\overline{%
{\rm {H}}}$ production as a function of $\gamma $. This is shown in figure 5
where the solid curve shows the cross section as given by eq. (4.1), and the
dashed line is the cross section with only the inclusion of $\sigma _{\gamma
^{*}T}.$ We see that for $\gamma $ of the order of 8, or larger, the cross
section is dominated by the transverse virtual photon component. Moreover we
also observe that the cross section for large $\gamma $'s disagree with the
equivalent photon approximation (EPA) used in ref. \cite{Mu94} to calculate
the cross section for the production of $\overline{{\rm {H}}}.$ It is thus
convenient to study under what circumstances the equivalent photon
approximation is reliable.

It might appear strange that the longitudinal part of the cross section is
relevant until such large values of $\gamma .$ It is thus reasonable to
check this result with a more schematic calculation. This can be achieved by
using plane-waves for the electron and hydrogenic waves for the positron,
without the correction terms to order $Z\alpha .$ In this case, we get

\begin{equation}
F({\bf Q})=4ie\sqrt{\pi }a_0^{-5/2}\;\frac{\overline{{\rm u}}(%
\overrightarrow{\gamma }{\bf .K}){\rm v}}{\left( {\bf Q-p}\right) ^2}\;,
\end{equation}
which yields the cross section

\begin{equation}
\frac{d\sigma }{d\varepsilon d\Omega }=\frac 1{\pi \omega }\;\left( \frac{%
Z_Te}{{\rm v}}\right) ^2\int_0^\infty d(q_t^2)\;\frac{q_t^2}{(Q^2-\omega
^2)^2}\;\Bigg[\frac{d\sigma _{\gamma ^{*}T}}{d\Omega }\left( {\bf Q},\omega
\right) +\frac{d\sigma _{\gamma ^{*}L}}{d\Omega }\left( {\bf Q},\omega
\right) \Bigg]\;,
\end{equation}
where the interference term is exactly zero, and

\begin{equation}
\frac{d\sigma _{\gamma ^{*}T}}{d\Omega }\left( {\bf Q},\omega \right) =\frac{%
32e^2p}{a_0^5}\frac 1{\left( {\bf Q-p}\right) ^8}\;,\;\;{\rm and}\;\;\frac{%
d\sigma _{\gamma ^{*}L}}{d\Omega }\left( {\bf Q},\omega \right) =\frac{%
32e^2p\omega ^2}{a_0^5q_t^2}\left( \frac 1{\gamma ^2{\rm v}}\right) ^2\frac 1%
{\left( {\bf Q-p}\right) ^8}\;.
\end{equation}

The integral over $\Omega $ in the expressions above can be done
analytically yielding

\begin{equation}
\left\{ 
\begin{tabular}{l}
$\sigma _{\gamma ^{*}T}\left( Q,\omega \right) $ \\ 
$\sigma _{\gamma ^{*}L}\left( Q,\omega \right) $%
\end{tabular}
\right\} =\frac{128\pi e^2p}{3a_0^5}\frac{\left( p^2+3Q^2\right) \left(
3p^2+Q^2\right) }{\left( Q^2-p^2\right) ^6}\left\{ 
\begin{tabular}{l}
$1$ \\ 
$\omega ^2/(q_t\gamma ^2{\rm v})^2$%
\end{tabular}
\right\} \;.
\end{equation}
Using this result, the integral over $q_t^2$ in (4.3) can also be performed
analytically but resulting in too long expressions to be transcribed here.
In figure 6 we plot $1-(d\sigma _L/d\varepsilon )/(d\sigma _T/d\varepsilon )$
obtained in this approximation, for $\varepsilon =1,$ 2, and 3,
respectively, and as a function of $\gamma .$ We observe indeed that even
for relatively large values of $\gamma $ (e.g., $\gamma \sim 5)$ the
longitudinal part of the cross section is still substantially relevant for
the calculation of the total cross section.

\section{The equivalent photon approximation}

The equivalent photon approximation is a well known method to obtain cross
sections for virtual photon processes in QED. It is described in several
textbooks (see, e.g., \cite{Be79}). It is valid for large $\gamma $'s, in
which case the cross section is dominated by transverse photons, as with the
production cross section of $\overline{{\rm {H}}}.$ In figure 7 we plot the
virtual photon cross section $\sigma _{\gamma ^{*}T}\left( Q,\omega \right) $
for $\varepsilon =1,$ $5,$ and 10, for $\gamma =100$, and as a function of $%
q_t.$ In this logarithmic plot it is evident that the function $\sigma
_{\gamma ^{*}T}\left( Q,\omega \right) $ is approximately constant until a
cutoff value, $q_t^{\max }$, at which it drops rapidly to zero. On the other
hand, the function

\begin{equation}
n\left( \gamma ,Q\right) =\frac 1\pi \;\left( \frac{Z_Te}{{\rm v}}\right) ^2%
\frac{\;q_t^2}{(Q^2-\omega ^2)^2}
\end{equation}
which we call by ``equivalent photon number'', is strongly peaked at $%
q_t\simeq \omega /\gamma {\rm v}$, which is very small for $\gamma \gg 1$.
Thus it is fair to write

\begin{equation}
\frac{d\sigma }{d\varepsilon }=\int_0^{q_t^{\max }}d\left( q_t^2\right) \;%
\frac 1\omega \;n\left( \gamma ,Q\right) \;\sigma _{\gamma ^{*}T}\left(
q_t=0,\omega \right) \simeq \frac 1\omega \;\sigma _\gamma \left( \omega
\right) \;\int_0^{q_t^{\max }}d\left( q_t^2\right) \;n\left( \gamma ,Q\right)
\end{equation}
where in the last equality we have approximated $\sigma _{\gamma
^{*}T}\left( q_t=0,\omega \right) \simeq \sigma _\gamma \left( \omega
\right) .$ This approximation is indeed valid for $\gamma \gg 1$, as we can
see from figure 8, where we plot the ratio between these two quantities for $%
\varepsilon =1,$ $5,$ and 10, and as a function of $\gamma $. For $\gamma $
bigger than 10 the approximation is quite good, and it is even better for
the lower values of $\varepsilon $. Eq. (5.2) is known as the ``equivalent
photon approximation''. The problem with the approximation is that the
integral in eq. (5.2) diverges logarithmically. The approximation is only
valid if we include a cutoff parameter $q_t^{\max }$, determined by the
value of $q_t$ at which $\sigma _{\gamma ^{*}T}\left( Q,\omega \right) $
drops to zero. A hint to obtain this parameter is to look at figure 7. We
see that the transverse virtual photon cross section drops to zero at $%
q_t\simeq \varepsilon .$ Another hint comes from figure 9, where we plot the
energy spectrum of the electron, $d\sigma /d\varepsilon $, for $\gamma
=3,\,\;10,\;$and $10$, respectively, obtained from a numerical integration
of eq. (4.1). We observe that the energy spectrum peaks at $\varepsilon
\simeq 2-3$, irrespective of the value of $\gamma $. We thus conclude that
an appropriate value of the cutoff parameter is $q_t^{\max }=2$. Inserting
this value in eq. (5.2) we get

\begin{equation}
\frac{d\sigma }{d\varepsilon }=\frac 1\pi \;\left( \frac{Z_Te}{{\rm v}}%
\right) ^2\frac 1\omega \;\sigma _\gamma \left( \omega \right) \;\left[ \ln
\left( 4x+1\right) -4\frac x{4x+1}\right] ,\;\;\;\;{\rm {where}\;\;\;\;}x=%
\frac{\gamma ^2-1}{\omega ^2}{\rm \;.}
\end{equation}

The dotted curve shown in figure 5 is the integral of the above equation
over $\omega $. We see that this approximation is very reasonable, for large 
$\gamma .$ It is only for the very extremely large $\gamma $'s that we
obtain $\sigma =(2.86$ pb) $Z_T^2\;\ln \gamma $, where the number inside
parenthesis comes from $(2\alpha /\pi )\int_2^\infty d\omega \;\sigma
_\gamma \left( \omega \right) /\omega =2.86$ pb. Only at $\gamma \sim 10^3$,
and larger, the approximation presented in ref. \cite{Mu94} becomes close
(in the range of 10\%) to the result obtained with the equation above.

\section{Conclusions}

We have calculated the cross sections for the production of antihydrogen in
collisions of antiprotons with heavy nuclear targets The cross section is
well described either by a semiclassical method or by the plane-wave Born
approximation. In fact, both approaches yield the same result for
relativistic antiprotons. The cross section can be separated into
longitudinal and transverse components, corresponding to the velocity of the
incident particle. For high energies, $\gamma \gg 1$, this corresponds to
the usual meaning longitudinal and transverse components of the virtual
photon. A very transparent and simple formulation is obtained using the
lowest order corrections to the positron and electron wavefunctions. At
ultra-relativistic antiproton energies, the contribution of the longitudinal
virtual photons to the total cross section vanishes. The remaining terms of
the cross section can be factorized in terms of a virtual photon spectra and
the cross section induced by real photons. However, the factorization
depends on a cutoff parameter, which is not well defined. This leads to
differences up to a factor of 2 between our results and those based on the
equivalent photon approximation. Production of antiatoms with the electron
at low energies, typically $\varepsilon \simeq 2$, is favored. The angular
distribution of the electrons is forward peaked within a angular interval $%
\Delta \theta \simeq 1/\varepsilon $.

The equivalent photon approximation is only of help as a qualitative
guidance for low values of $\gamma $ ($\gamma $ smaller than about 10). This
is the case for the CERN/LEAR experiment ($p=1.94$GeV/c), to a lesser extent
for the Fermilab experiment \cite{Fe97} where the beam momentum can vary
from 3.45 to 8.8 GeV/c. The formulation presented here allows for a better
quantitative calculation of the production of antiatoms at these energies.
The discrepancy to earlier calculations \cite{Be87} for small values of $%
\gamma$ is at present not understood.

\bigskip

{\Large {\bf Acknowledgments}}

This work was partially supported by the KFA-Juelich, and by the
MCT/FINEP/CNPQ(PRONEX) under contract No. 41.96.0886.00.

\appendix

\section{Positron wavefunction}

Here we deduce a first order ($Z\alpha $) correction of the positron
wavefunction, important in the calculation of the $\overline{H}$ production
cross section. To be general, we consider the production of a bound positron
in a relativistic antinucleus, with charge $-Z$, incident on a target
nucleus with charge $Z_T$. Much of our calculation is based on arguments
presented in ref. \cite{Be79} with connection to the photoelectric effect.

The Dirac equation for a positron in the field of the antinucleus is:

\begin{equation}
\varepsilon \Psi =\left[ \overrightarrow{\alpha }{\bf .}({\bf p}+e{\bf A}%
)-\beta m-e\phi \right] \Psi \;.
\end{equation}

For ${\bf A}=0:$

\begin{equation}
\left[ \varepsilon -U+\beta m+i\overrightarrow{\alpha }{\bf .}%
\overrightarrow{\nabla }\right] \Psi =0\;,
\end{equation}
where $U=-e\phi =-Ze^2/r$ is the Coulomb field. The positron will be most
likely be produced at the s-state of the antinucleus. To lowest order the
wavefunction is given by the non-relativistic hydrogenic wavefunction

\begin{equation}
\Psi _{non-r}(r)=\frac 1{\sqrt{\pi }}\;\left( \frac 1{a_0}\right)
^{3/2}\;e^{-r/a_0}\;,
\end{equation}
where $a_0=1/(Ze^2m)$.

To first-order, a corrected wavefunction (to order $Ze^2$) is given by

\begin{equation}
\Psi ={\rm {v}}\Psi _{non-r}+\Psi ^{(1)}\;,
\end{equation}
where {\rm v} denotes the positron spinor.

Applying the operator $\varepsilon -U-m\beta +i\overrightarrow{\alpha }{\bf .%
}\overrightarrow{\nabla }$ to (A.2) we get

\begin{equation}
\left( \nabla ^2+p^2-2\varepsilon U\right) \Psi =\left( i\overrightarrow{%
\alpha }{\bf .}\overrightarrow{\nabla }U-U^2\right) \Psi .
\end{equation}

Substituting (A.4) into (A.5) and expanding:

\begin{eqnarray}
&&\left( \nabla ^2+p^2-2\varepsilon U\right) \Psi _{non-r}{\rm {v}}+\left(
\nabla ^2+p^2-2\varepsilon U\right) \Psi ^{(1)}  \nonumber \\
&& \\
&=&\;\left( i\overrightarrow{\alpha }{\bf .}\overrightarrow{\nabla }%
U-U^2\right) \Psi _{non-r}{\rm {v}}+\left( i\overrightarrow{\alpha }{\bf .}%
\overrightarrow{\nabla }U-U^2\right) \Psi ^{(1)}.  \nonumber
\end{eqnarray}

Since $p^2=\varepsilon ^2+m^2\simeq -2m|\varepsilon _s|$, $\Psi
^{(1)}\propto Ze^2$, and $U\propto Ze^2$, we get to lowest order 
\[
\left( \nabla ^2+p^2-2\varepsilon U\right) \Psi _{non-r}\simeq \left( \nabla
^2-2m|\varepsilon _s|-2|\varepsilon _s|U\right) \Psi _{non-r}\simeq 0 
\]
and eq. (A.6) becomes

\[
(\nabla ^2-2m|\varepsilon _s|-2mU)\Psi ^{(1)}=\left( i\overrightarrow{\alpha 
}{\bf .}\overrightarrow{\nabla }U\right) \Psi _{non-r}{\,}{\rm {v}}\;, 
\]

or

\begin{equation}
(\frac{\nabla ^2}{2m}-|\varepsilon _s|-\frac{Ze^2}r)\Psi ^{(1)}=-\frac{i%
\overrightarrow{\alpha }}{2m}{\bf .}\left( \overrightarrow{\nabla }\frac{Ze^2%
}r\right) \Psi _{non-r}{v}\;,
\end{equation}

The non-relativistic wave function obeys the equation $(\nabla
^2/2m-|\varepsilon _s|+Ze^2/r)\Psi _{non-r}=0$, from which we deduce that

\begin{equation}
(\frac{\nabla ^2}{2m}-|\varepsilon _s|-\frac{Ze^2}r)\overrightarrow{\nabla }%
\Psi _{non-r}=-\left( \overrightarrow{\nabla }\frac{Ze^2}r\right) \Psi
_{non-r}\;,
\end{equation}

Thus, if $\Psi ^{(1)}=i/(2m)${\rm v}$\overrightarrow{\alpha }{\bf .}%
\overrightarrow{\nabla }\Psi _{non-r}$, it will be the solution of (A.7). An
approximate solution of (A.1) is therefore

\begin{equation}
\Psi ^{(+)}=\left[ 1+\frac{i\overrightarrow{\alpha }}{2m}{\bf .}%
\overrightarrow{\nabla }\right] {v}\Psi _{non-r}=\left[ 1+\frac i{2m}%
\;\gamma ^0\overrightarrow{\gamma }{\bf .}\overrightarrow{\nabla }\right] 
{\rm {v}}\Psi _{non-r}\;,
\end{equation}

The relevant distances for the non-relativistic wavefunction are $r\sim
1/(mZe^2)$. The correction term should be good within these distances. But,
for the ground state (or any s-state) it can be used for any value of $r$,
since the derivative of the exponential function (A.3) is always
proportional to $Ze^2.$ Because of that, we can use the corrected
wavefunction in our calculation of $\overline{H}$ production where, as we
see in section 2, the small values of r are essential in the computation of
the matrix elements.

\section{Electron wavefunction}

For the electron wavefunction we use a plane wave and a correction term to
account for the distortion due to the antinucleus charge. As in Appendix A,
the correction term is considered to be proportional to $Ze^2$. The
wavefunction is then given by

\begin{equation}
\Psi ={\rm u\ }e^{i{\bf p.r}}+\Psi ^{\prime }\;.
\end{equation}

In section 2 we show that only the Fourier transform of $\Psi ^{\prime }$
will enter the calculation. This Fourier transform can be deduced directly
from the Dirac equation for the electron in the presence of a Coulomb field
of an antinucleus:

\begin{equation}
\left( \gamma ^0\varepsilon +i\overrightarrow{\gamma }{\bf .}\overrightarrow{%
\nabla }{\bf -}m\right) \Psi ^{\prime }=\frac{Ze^2}r\;\gamma ^0{\rm u}\;e^{i%
{\bf p.r}},
\end{equation}

Applying on both sides of this equation the operator $\left( \gamma
^0\varepsilon +i\overrightarrow{\gamma }{\bf .}\overrightarrow{\nabla }{\bf +%
}m\right) $ we get

\begin{equation}
\left( \Delta +p^2\right) \Psi ^{\prime }=Ze^2\left( \gamma ^0\varepsilon +i%
\overrightarrow{\gamma }{\bf .}\overrightarrow{\nabla }{\bf +}m\right)
(\gamma ^0{\rm u})\frac{e^{i{\bf p.r}}}r\;.
\end{equation}

Multiplying by $e^{-i{\bf q.r}}\;$and integrating over $d^3r$ we get

\begin{equation}
\left( p^2-q^2\right) \Psi _{{\bf q}}^{\prime }=Ze^2\left[ 2\gamma
^0\varepsilon +i\overrightarrow{\gamma }{\bf .(q-p)}\right] (\gamma ^0{\rm u}%
)\frac{4\pi }{({\bf q-p})^2}\;,
\end{equation}

where we have used the identity $\left( \gamma ^0\varepsilon +i%
\overrightarrow{\gamma }{\bf .}\overrightarrow{\nabla }{\bf -}m\right)
(\gamma ^0{\rm u})=0.$ Thus,

\begin{equation}
\Psi _{{\bf q}}^{\prime }\equiv \left( \Psi _{{\bf q}}^{\prime }\right)
^{*}\gamma ^0=-4\pi Ze^2\overline{{\rm u}}\frac{2\gamma ^0\varepsilon +i%
\overrightarrow{\gamma }{\bf .(q-p)}}{({\bf q-p})^2(q^2-p^2)}
\end{equation}

In section 2 we use this equation to calculate the matrix element for the
production of anti-atom.

\section{Plane-wave Born approximation}

In the plane wave Born approximation the transition matrix element is given
by

\begin{equation}
T_{fi}=\int d^3r\;\left[ \rho _{fi}\left( {\bf r}\right) \phi ({\bf r})-{\bf %
j}_{fi}{\bf (r).A}({\bf r})\right]
\end{equation}
where

\begin{equation}
\left\{ 
\begin{array}{c}
\phi ({\bf r}) \\ 
{\bf A}({\bf r})
\end{array}
\right\} =\left\{ 
\begin{array}{c}
1 \\ 
{\bf v}
\end{array}
\right\} 
\displaystyle \int 
d^3r^{\prime }\;\frac{e^{i\omega \left| {\bf r-r}^{\prime }\right| }}{\left| 
{\bf r-r}^{\prime }\right| }\;\left\langle {\bf k}_f|{\bf r}^{\prime
}\right\rangle \left\langle {\bf r}^{\prime }|{\bf k}_i\right\rangle \;,
\end{equation}
and $\left\langle {\bf k|r}\right\rangle =e^{i{\bf k.r}}$ is a plane wave
for the antiproton. Using

\begin{equation}
\frac{e^{i\omega \left| {\bf r-r}^{\prime }\right| }}{\left| {\bf r-r}%
^{\prime }\right| }=\frac 1{2\pi ^2}%
\displaystyle \int 
d^3K\;\frac{e^{i{\bf K.(r-r}^{\prime }{\bf )}}}{K^2-\omega ^2}
\end{equation}
and ${\bf J}_{fi}={\bf v}\rho _{fi},$ we get

\begin{equation}
T_{fi}=4\pi \int d^3r\;\frac{e^{i{\bf Q.r}}}{Q^2-\omega ^2}\;\left[ \rho
_{fi}\left( {\bf r}\right) -{\bf j}_{fi}{\bf (r).v}\right] \;=4\pi Ze\;\frac{%
F({\bf Q})}{Q^2-\omega ^2},
\end{equation}
where ${\bf Q=k}_i{\bf -k}_f$ and $F({\bf Q})$ is given by eq. (2.6). The
cross section is given by

\[
\frac{d\sigma }{d\Omega }=\left( \frac E{2\pi }\right) ^2\sum_{spins}%
\displaystyle \int 
\;\left| T_{fi}\right| ^2\;. 
\]

For relativistic antiproton energies

\begin{eqnarray}
Q_L &=&k_i-k_f\cos \theta \simeq k_i-k_f\simeq \omega /{\rm v}  \nonumber \\
Q_T &\equiv &q_t=k_f\sin \left( \theta \right) \simeq Ev\sin \left( \theta
\right) \;\;\;\;\Longrightarrow \;\;\;\;d\Omega =d^2q_t/\left( E{\rm v}%
\right) ^2\;,
\end{eqnarray}
so that

\begin{equation}
\sigma =4\left( \frac{Z_Te}{{\rm v}}\right) ^2\sum_{spins}%
\displaystyle \int 
d^2q_t\;\frac{\left| F({\bf Q})\right| ^2}{\Big(Q^2-\omega^2\Big)^2}\;.
\end{equation}

The $q_t$ integration ranges from 0 to a maximum value $E{\rm v}$, where $E$
and ${\rm v}$ are the antiproton energy and velocity, respectively. This
value is however much larger than the relevant energies entering the matrix
elements in $F({\bf Q})$. Thus, the expression above is the same as the one
derived in section 2, eq. (2.7).

\section{Figure captions}

\begin{enumerate}
\item  {} Diagram for $\overline{{\rm H}}$ in collisions of $\overline{{\rm p%
}}$ with a nuclear target.

\item  {} Angular distribution of the electrons in $\overline{{\rm H}}$
production by real photons for several electron energies.

\item  {} Production cross sections by longitudinal and transverse virtual
photons, $\sigma _{\gamma ^{*}T}$, $\sigma _{\gamma ^{*}L}$, respectively.
The cross sections are calculated for $\varepsilon =2,$ and $\gamma =3$.

\item  {} Same as in figure 3, but for $\gamma =10$ and $\varepsilon =2$.

\item  {} $\overline{{\rm H}}$ production in a collision of an antiproton
with a proton target, as a function of $\gamma $. The solid curve shows the
cross section as given by eq. (4.1), and the dashed line is the cross
section with only the inclusion of $\sigma _{\gamma ^{*}T}$. The dashed line
is the result of the approximation given by eq. (5.3).

\item  $\overline{{\rm H}}$ production in a collision of an antiproton with
a proton target calculated using a plane-wavefunction for the electron and a
hydrogenic wavefunction for $\overline{{\rm H}}$. The ratio $(d\sigma
_T/d\varepsilon -d\sigma _L/d\varepsilon )/d\sigma _T/d\varepsilon $ is
shown as a function of $\gamma $ and for $\varepsilon =1,\;2,\;$ and 3,
respectively. Eqs. (4.3 - 4.5) were used in the calculation.

\item  {} Virtual photon cross section $\sigma _{\gamma ^{*}T}\left(
Q,\omega \right) $ for $\varepsilon =1,$ $5,$ and 10, for $\gamma =100$, and
as a function of $q_t.$

\item  {} Ratio between $\sigma _{\gamma ^{*}T}\left( q_t=0,\omega \right) $
and $\sigma _\gamma \left( \omega \right) $ for $\varepsilon =1,$ $5,$ and
10, and as a function of $\gamma $.

\item  {} Energy spectrum of the electron, $d\sigma /d\varepsilon $, for $%
\gamma =3,\,\;10,\;$and $10$, respectively, obtained from a numerical
integration of eq. (4.1).
\end{enumerate}


\bigskip

\bigskip

\begin{references}
\bibitem{Ba96}  PS210 collaboration,W. Oelert, spokesperson; G. Baur et al.,
Phys. Lett. {\bf B368 }(1996) 251

\bibitem{Mu94}  C.T. Munger, S.J. Brodsky, and I. Schmidt, Phys. Rev. {\bf %
D49 }(1994) 3228

\bibitem{BB94}  C.A. Bertulani and G. Baur, Physics Today, March 1994, p.
22; and further references given there.

\bibitem{Be88}  C.A. Bertulani and G. Baur, Phys. Rep. {\bf 163 }(1988) 299;
Nucl. Phys. {\bf A505 }(1989) 835

\bibitem{As94}  A. Aste et al., Phys. Rev. {\bf A50 }(1994) 3980

\bibitem{Ba93}  A.J. Baltz et al., Phys. Rev {\bf A48 }(1993) 2002

\bibitem{Bau93}  G. Baur, Phys. Lett. {\bf B311 }(1993) 343

\bibitem{Fe97}  M. Mandelkern, letter of intent/experiment E862, Fermilab,
available at http://fnphyx-www.fnal.gov/experiments/e862/e862.html

\bibitem{Be79}  V.B. Berestetskii, E.M. Lifshitz, and L.P. Pitaevskii, {\it %
Relativistic Quantum Field Theory}, {\it Part 1}, 2nd ed. (Pergamon, New
York, 1979)

\bibitem{He54}  W. Heitler, {\it The Quantum Theory of Radiation}, 3rd. ed.
(Oxford, New York, 1954)

\bibitem{err1}  The eq. (33.16) of ref. \cite{Be79} has a wrong sign in the
last term. The correct result, originally due to Sauter \cite{Sau}, is
presented in the book of Heitler \cite{He54}

\bibitem{Sau}  F. Sauter, Ann. Physik {\bf 9} (1931) 217; ibid. {\bf 11}
(1931) 454

\bibitem{Be87}  U. Becker, J. Phys. {\bf B20} (1987) 6563
\end{references}
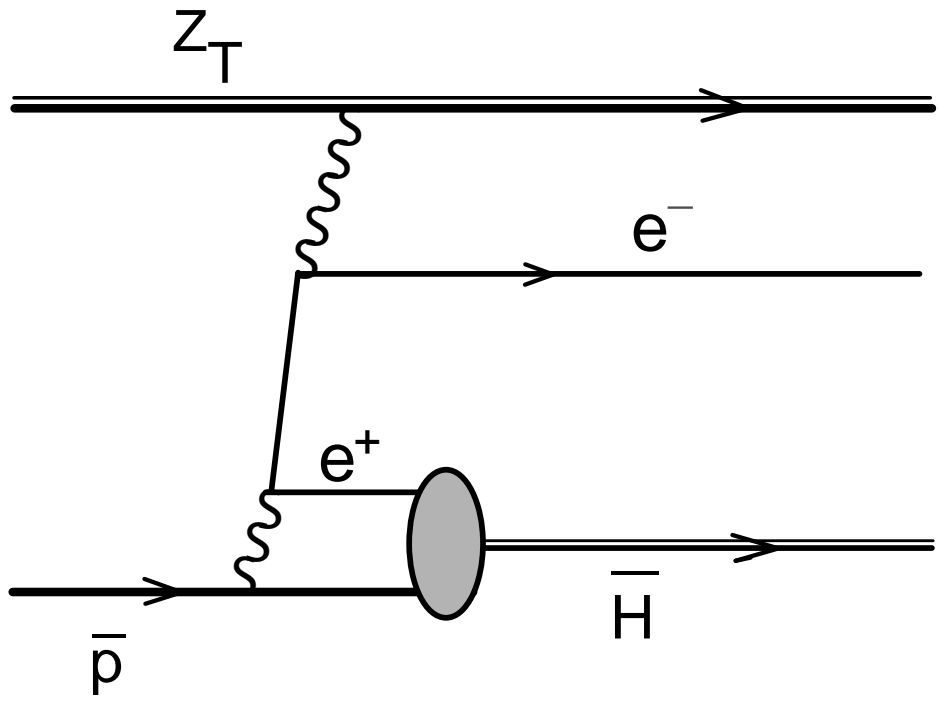
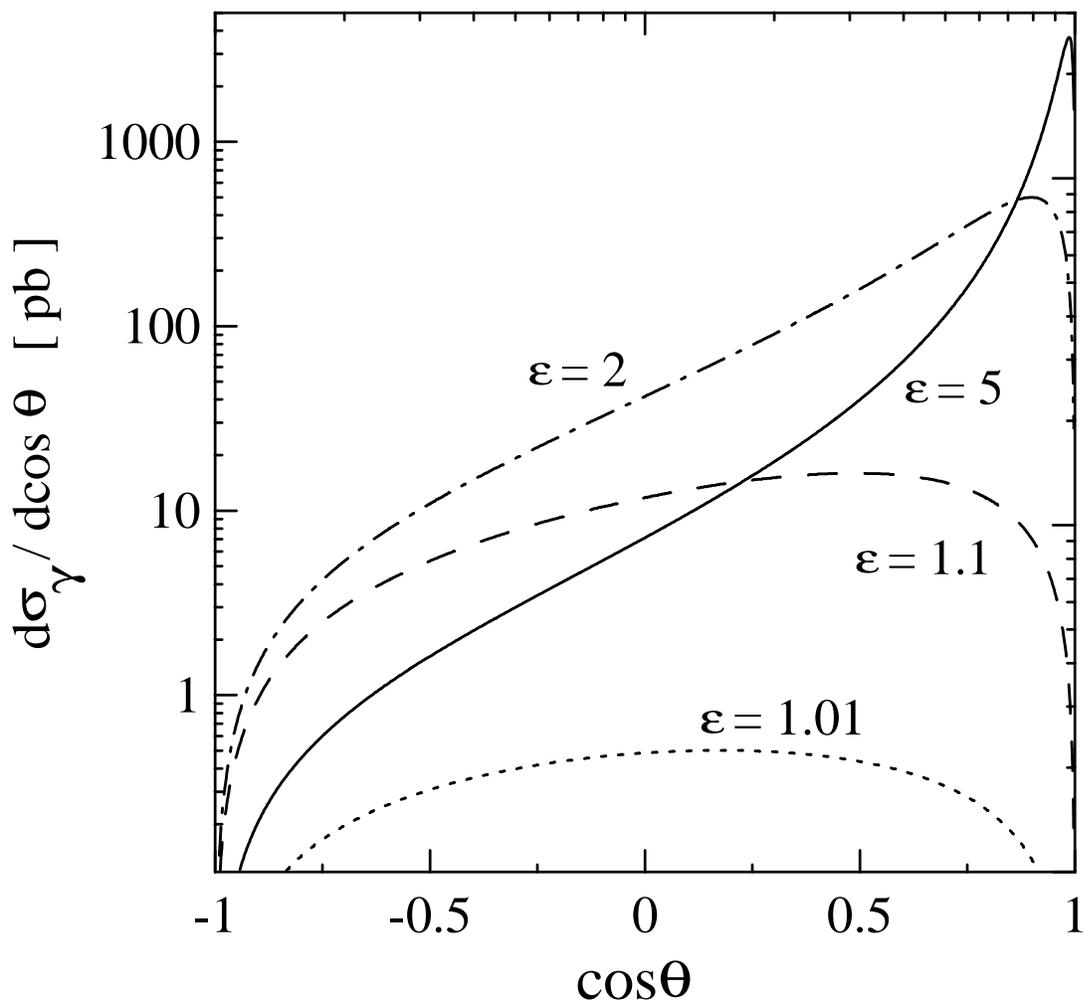
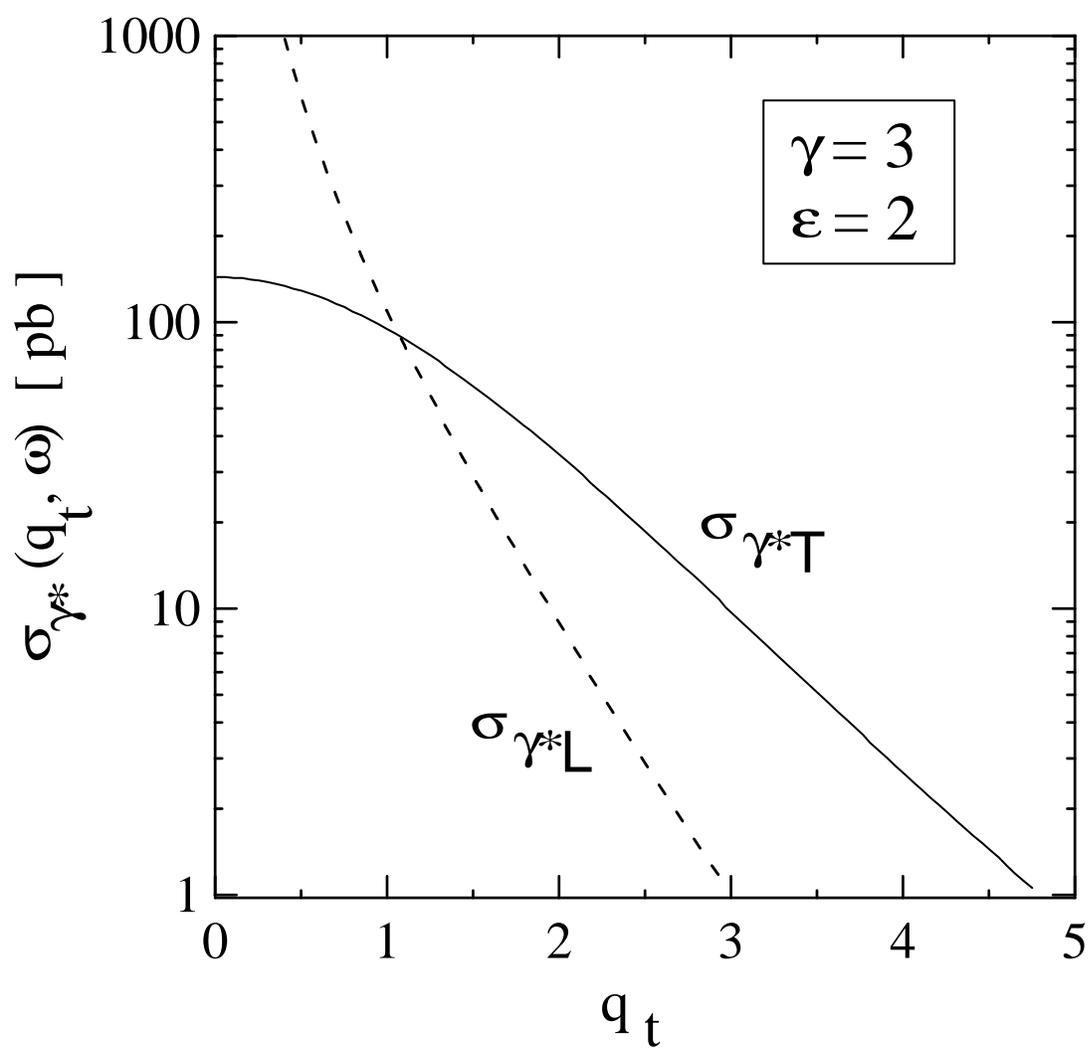
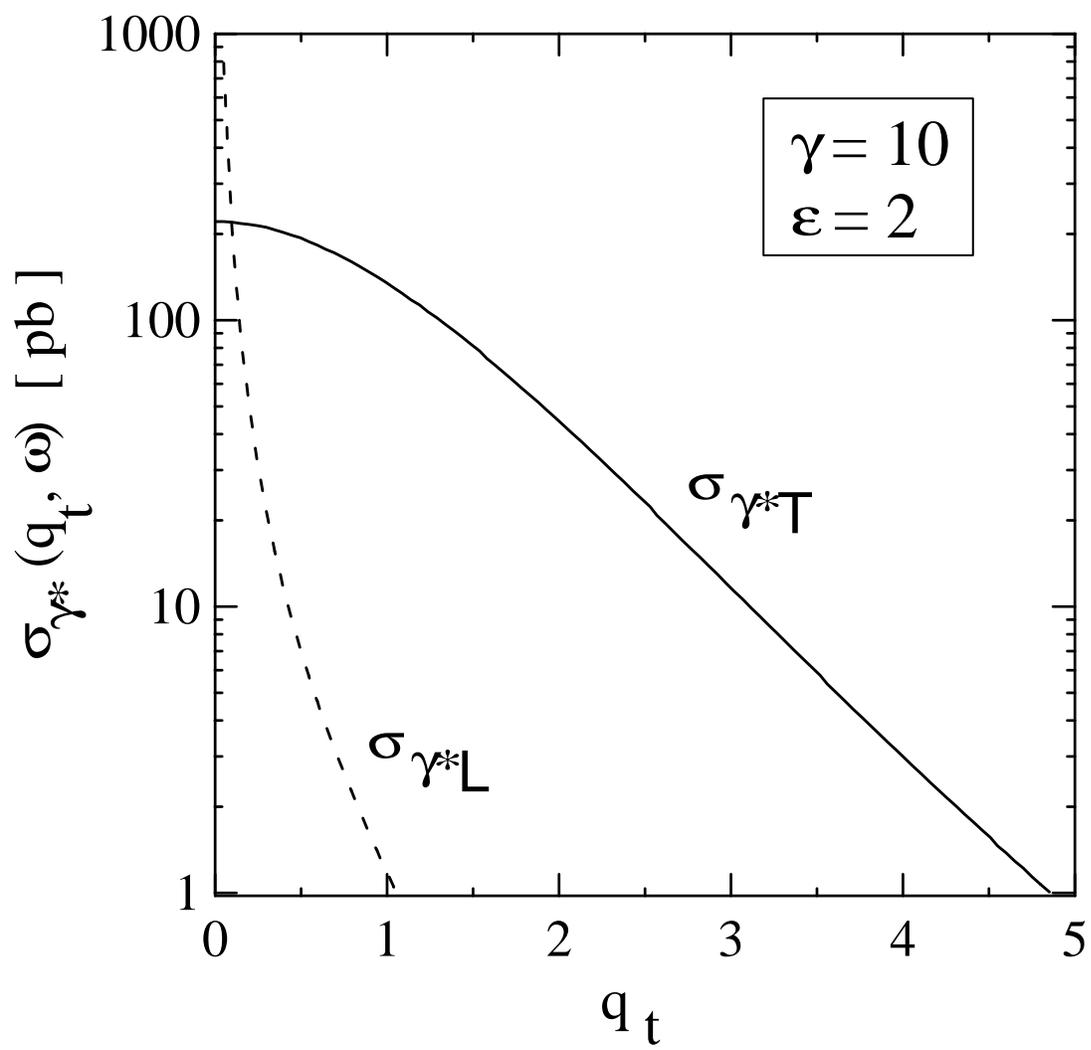
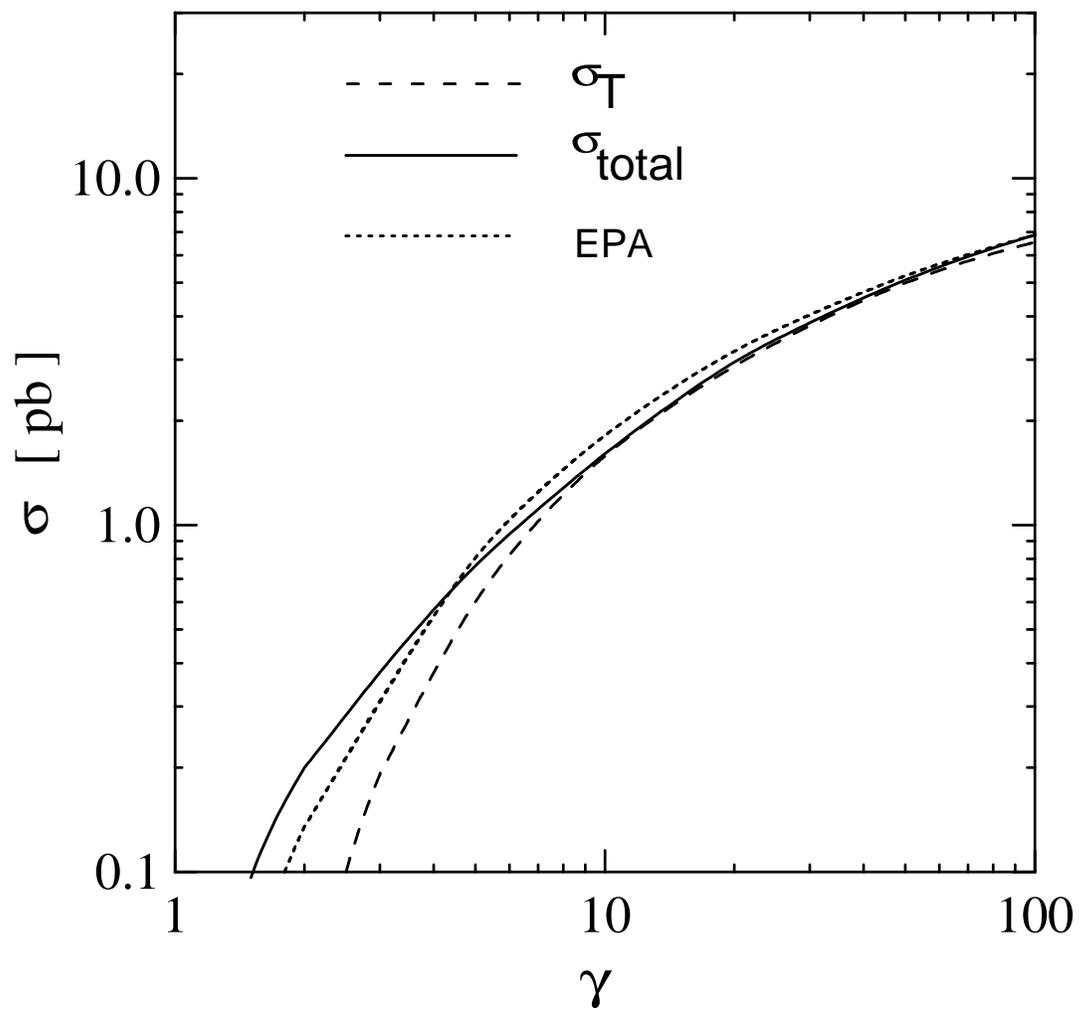
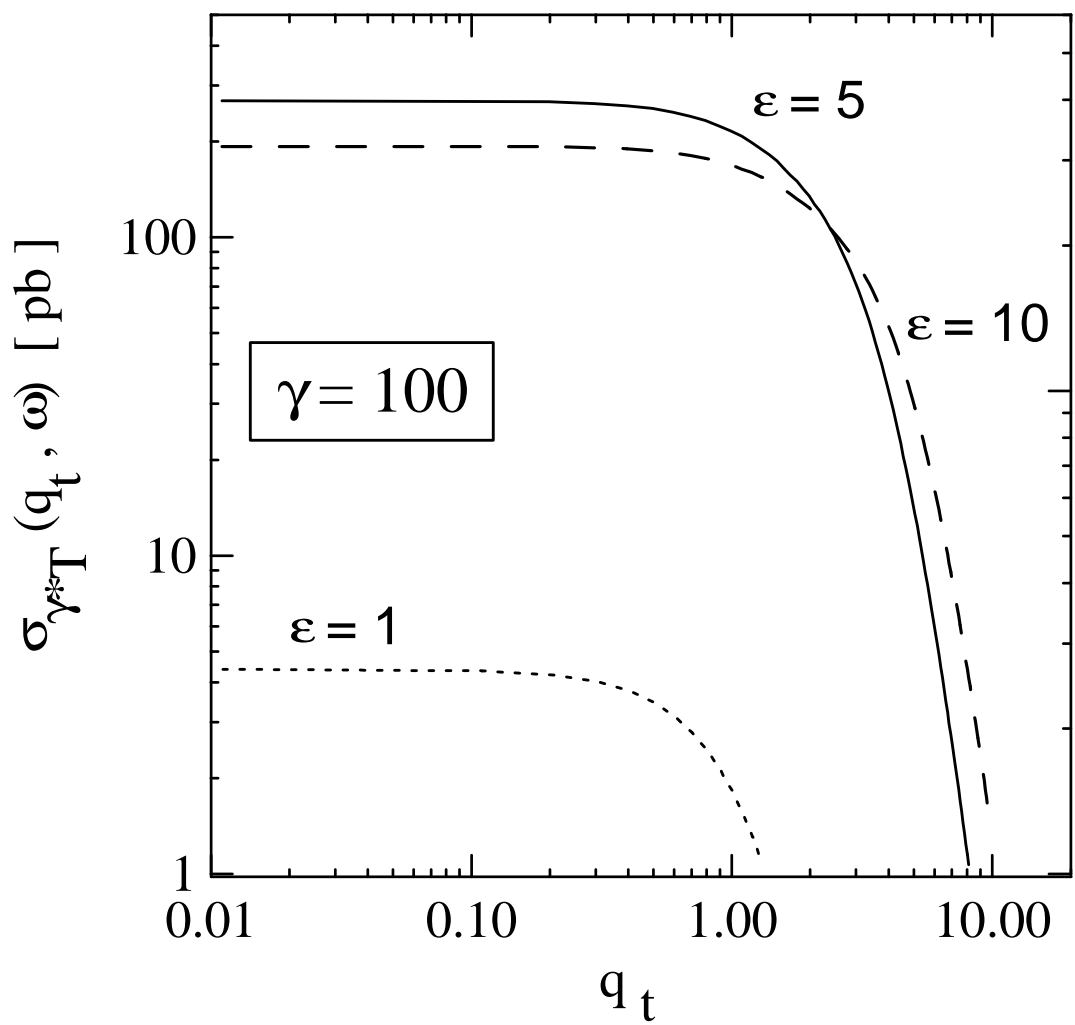
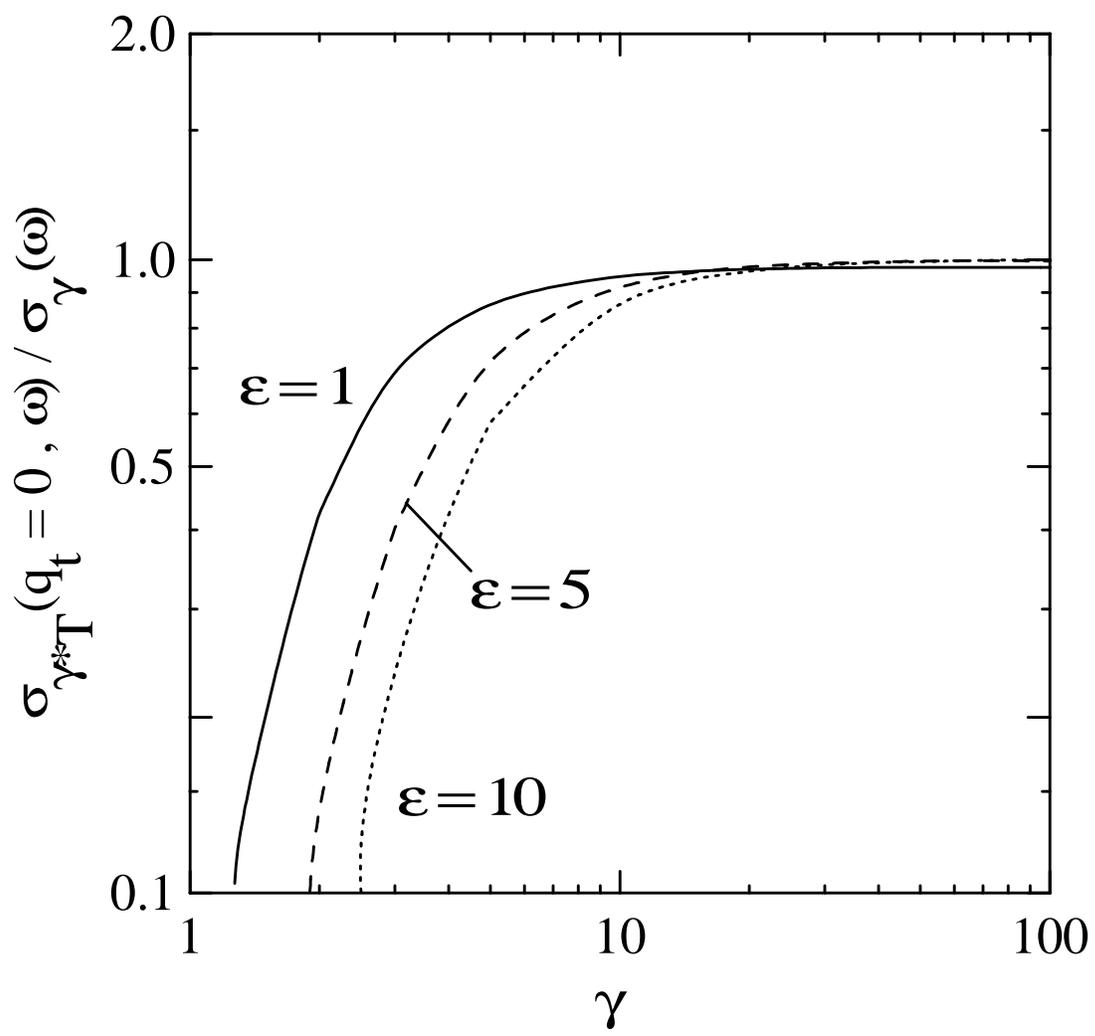
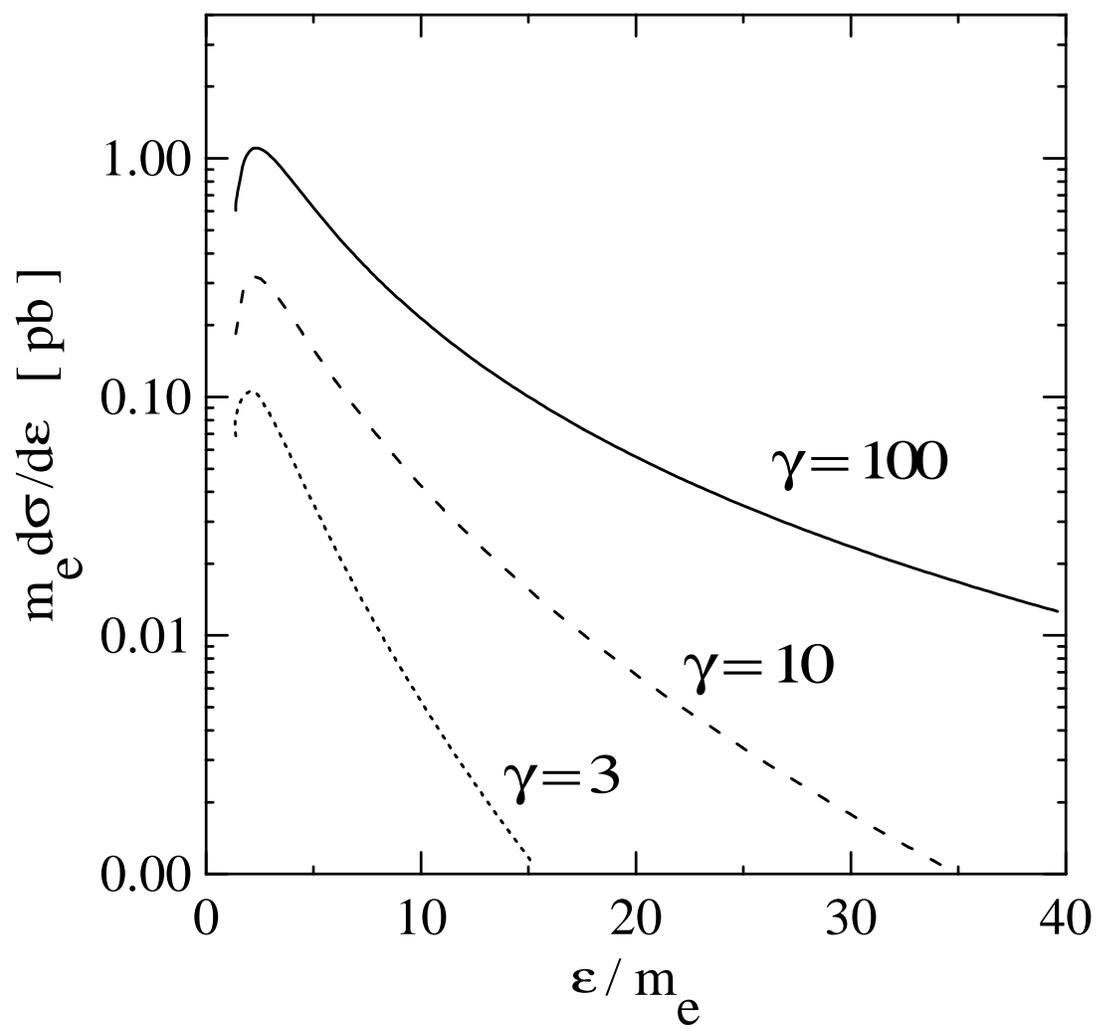
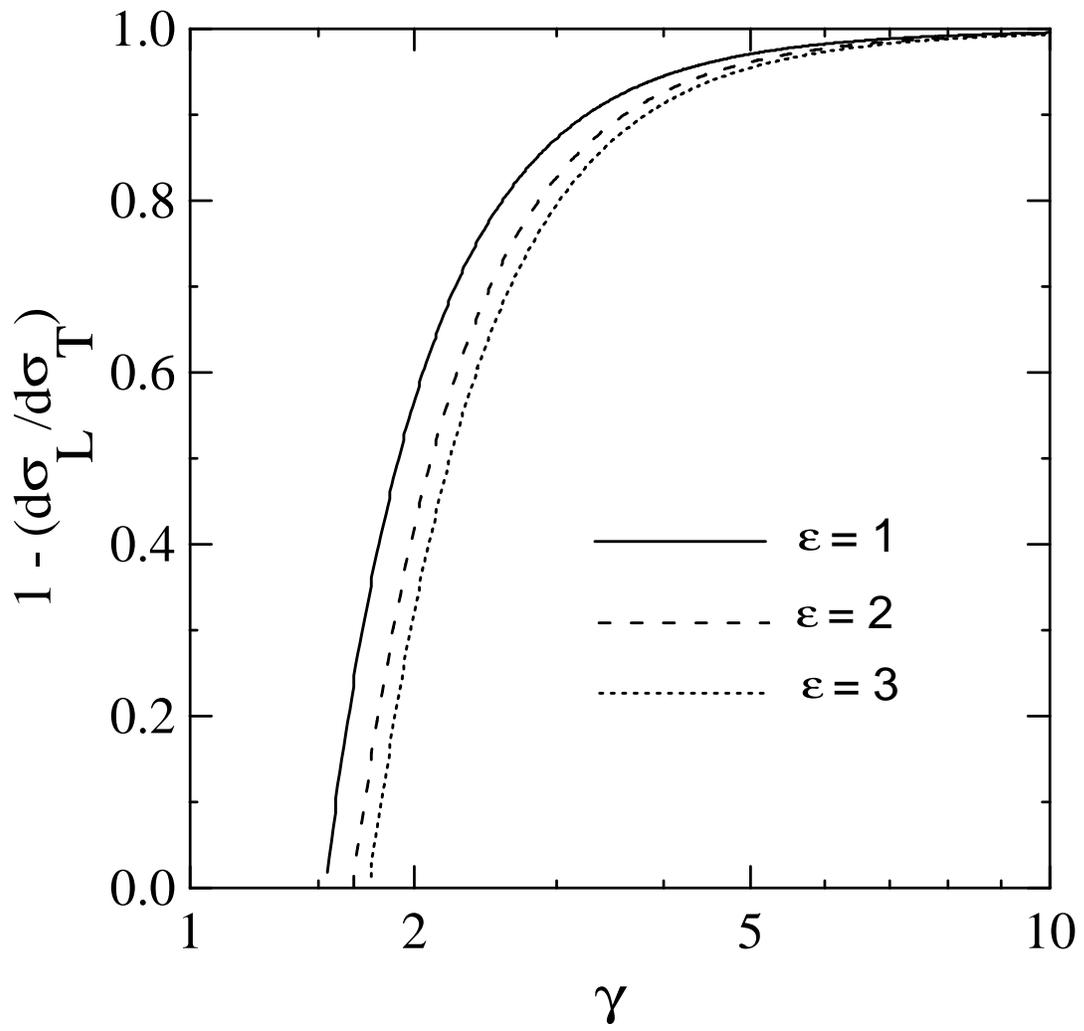
\end{document}